\documentclass[usenatbib]{mn2e}
\usepackage{float}
\usepackage{graphicx}
\usepackage{amsmath}
\usepackage{placeins}
\usepackage{threeparttable}
\usepackage{mathptmx}
\usepackage{txfonts}
\usepackage[T1]{fontenc}
\usepackage{ae,aecompl}
\usepackage{pdfpages}
\usepackage{etex}
\usepackage{tablefootnote}
\usepackage{scalerel}
\usepackage{longtable}
\usepackage{hyperref}
\usepackage{amsmath,amssymb}
\hypersetup{colorlinks, citecolor=blue}
\newcommand\HI{H\protect\scaleto{$I$}{1.2ex}}
\newcommand{\kms}{\mbox{km\,s$^{-1}$}}
\begin{document}

\title[Stability of galaxies across morphological sequence]
{Stability of galaxies across morphological sequence}
\author[ K. Aditya]
        {K. Aditya \thanks{E-mail : aditya.k@iiap.res.in} \\
        Indian Institute of Astrophysics, Koramangala, Bengaluru 560 034, INDIA \\ 
} 
\maketitle 


\begin{abstract}
We investigate the stability of nearby disc galaxies and galaxies at redshift ($z$) equal to 4.5. We explore the connection between the stability parameter $(Q_{RW})$, star formation rate ($SFR$), gas fraction $(f^{Gas})$, and the time scale for growth of gravitational instabilities $(\tau)$. We find that, despite differences in morphology $91$ $\%$ of the nearby galaxies have a minimum value of stability parameter ($Q^{Min}_{RW}$) greater than $1$ indicating stability against the growth of axisymmetric instabilities. The spirals in our sample have higher median star formation rate, lower median $Q_{RW}$, a lower $f^{Gas}$ and  small time scale for growth of  gravitational instabilities than irregular galaxies.  We find that the gravitational instabilities in spirals convert a large fraction of gas into stars quickly, depleting the gas reservoirs.  On the other hand, star formation occurs more gradually over longer timescales in irregulars with a higher gas fraction. We then compare the stability of the nearby galaxies with galaxies at $z\,=\,4.5$.
We find that net stability levels in the nearby galaxies and the galaxies at $z\,=\,4.5$ are primarily driven by the stellar disc 
suggesting the presence of an inherent mechanism that self-regulates the stability. Finally, upon removing the contribution of the dark matter to the total potential, the median $Q_{RW}$ for the nearby galaxies and galaxies at $z \,= \,4.5$ remains unchanged indicating that the baryons can self-regulate the stability levels, at least in a statistical sense.
\end{abstract}
\section{Introduction}
Observations of galaxies reveal a wide range of morphological features and physical properties, which are encoded into
Hubble's classification scheme of galaxies. Different physical processes like tidal interaction, galaxy mergers,
fragmentation of the gas disc, turbulence, and feedback play an important role in shaping the observed morphology of the galaxies.
The tidal perturbations due to a passing satellite galaxy can alter the shape of both the target galaxy and the satellite leading to the
formation of tidal tails and spiral arms \citep{toomre1972galactic}. Also, the tidal perturbation arising from the high-speed head-on collision
of a companion galaxy with a target disc galaxy is responsible for ring formation and enhancement of star formation activity \citep{renaud2018morphology}.
Apart from direct tidal encounters, major and minor mergers can transform the spiral galaxies into elliptical galaxies
\citep{barnes1992formation,mihos1994triggering,bournaud2007multiple,pedrosa2014morphology}. Further, the transformation of the disc galaxies
into elliptical galaxies is accompanied by the quenching of the star
formation activity \citep{martig2009morphological}. \cite{scannapieco2009formation} consider the effect of various physical
processes like feedback and cooling of the gas and multi-phase treatment of interstellar medium (ISM) in hydrodynamical simulations. They find that the survival of disc galaxies in the $\Lambda$CDM
scenario is strongly affected by major mergers and disc instabilities. \cite{mihos1995merger} using numerical
simulations showed that mergers of disc galaxies with low-mass satellites can trigger the formation of strong bars
and also give rise to bending instabilities.\\
High-resolution observation of the ISM reveals that ISM is
inherently turbulent \citep{leroy2016portrait, brunetti2021highly, leroy2021phangs, meidt2023phangs}. \cite{bacchini2020evidence} show that the feedback from the
supernova explosions predominantly drive the turbulence observed in the ISM. \cite{agertz2015characterizing} study hydro + N-body simulations by including stellar
feedback, which leads to driven ISM turbulence. They find that models without stellar feedback show complete disc fragmentation, whereas the feedback-driven models
are stable on all scales. The disc fragmentation due to gravitational instabilities is consistent with the earlier theoretical formulation of \cite{goldreich1965gravitational}.
They consider the stability of the stratified self-gravitating sheet and show that the instabilities can break up the sheet into masses of definite sizes
if $\frac{\pi G \rho}{4 \omega^{2}}$ is greater than a number between 0.7 and 1.8, $\omega$ is the angular velocity, and $\rho$ is the density of the medium.
Furthermore, using the  JWST observations of nearby galaxies \cite{meidt2023phangs} have shown that the filamentary structures observed in the nearby
galaxies are consistent with the models of disc fragmentation driven by Jean's instabilities. \\
Studies by \cite{wang1994gravitational} and \cite{pandey1999gravitational} have shown that axisymmetric gravitational instabilities primarily drive
star formation activity. \cite{li2005control} show that the formation of the star clusters in numerical simulations results from axisymmetric
gravitational instability quantified by Toomre Q parameter \citep{toomre1964gravitational}. Also, observationally, it has been seen that the young stellar
objects increase with decreasing  Toomre Q \citep{gruendl2009high,dawson2013supergiant}. Toomre instabilities typically are 2D instabilities, but a recent study by
\cite{meidt2022molecular} shows that the fragmentation of the molecular clouds is brought about by 3D instabilities. Gravitational instabilities are thus at the heart of different physical processes that dictate the observed structure, morphology, and physical properties of the galaxies.\\
The observed physical properties of galaxies, i.e., the surface brightness profile $\Sigma$, the epicyclic motion $\kappa$, and radial velocity dispersion $\sigma$,
constitute essential ingredients for quantifying the stability/instability levels of the disc galaxies against the growth of
axisymmetric gravitational instabilities. The condition for quantifying the stability of a galaxy against the growth of axisymmetric instabilities
derived by \cite{toomre1964gravitational} is written as
\begin{equation}
Q=\frac{\kappa \sigma}{\pi G \Sigma}.
\end{equation}
A value of $Q>1$ indicates that the galaxy is stable against the growth of axisymmetric instabilities. The $'Q'$ parameter proposed by Toomre has been modified by \cite{jog1996local,rafikov2001local,romeo2011effective,romeo2013simple} to include the effect of multiple self-gravitating components. The stability parameter has also been
further modified to include the physical processes like the effects of the turbulence \citep{hoffmann2012effect, agertz2015characterizing} and the three-dimensional
structure of ISM \citep{meidt2022molecular}.
The multi-component stability parameter, apart from accounting for the self-gravity of the multiple mass components, also allows us to discern if 
the gravitational instabilities are driven by stars or by gas. The $'Q'$ parameter derived by \cite{jog1996local,rafikov2001local,romeo2011effective,romeo2013simple}
provides a simple method to quantify the stability levels in the galaxies. \cite{romeo2017drives} analyze the stability of a sample of $12$ nearby starforming
galaxies from the THINGS \citep{leroy2008star} using the multi-component stability parameter $(Q_{RW})$ derived by \cite{romeo2013simple}. They find that the median 
value of $Q_{RW}$ for the galaxies lies between $2$ and $3$, with a global median equal to $2.2$. Also, using a sample of $34$ spiral galaxies \cite{romeo2018angular}
point out that the mass-weighted average value of the stellar stability parameter $< Q_{*} >$ is constant across spiral galaxies of all morphological types (Sa - Sd).  \\
In this work, we will explore the role of gravitational instabilities in galaxies with diverse physical properties and morphologies by addressing the following questions:
\begin{enumerate}

\item How does the stability levels quantified by the two-component stability parameter $Q_{RW}$ vary as a function of the radius and morphology ?\\

\item Out of the stars and the gas which is dominant mass component driving the stability levels in the galaxies ?\\

\item What is the minimum value of stabilty parameter ($Q^{Min}_{RW}$) for the nearby galaxies ? Are the nearby galaxies stable against non-axisymmetric instabilities and gas dissipation ?\\

\item How does the star formation rates in galaxies which are susceptible to the growth of axisymmetric instabilities ($Q^{Min}_{RW}<1$) compare with galaxies which are stable against the growth of axisymmetric instabilities($Q^{Min}_{RW}>1$) ?\\

\item Where does the galaxy become unstable, i.e., attain a minimum value of $Q_{RW}$? Are the most unstable regions driven by stars or by gas ? \\

\item How does the stability in nearby galaxies compare with the galaxies observed at high redshift $(z=4.5)$ ?\\

\item What is the role of dark matter in regulating stability? Can baryons self-regulate the stability ?\\

\item What is the connection between $Q_{RW}$, gas fraction $(f^{Gas})$ and star formation rate $(SFR)$ with the time scale for growth of instabilities  $(\tau)$?
\end{enumerate}
In order to answer the above questions, we calculate the stability of a sample of $175$ nearby galaxies with diverse morphologies and physical properties from the SPARC galaxy catalog \citep{lelli2016sparc} using the two-component stability criterion $(Q_{RW})$ derived by \cite{romeo2011effective}. The two-component stability criterion considers the self-gravity of the stars and the gas on an equal footing and includes a correction for the disc thickness. The galaxies in the SPARC catalog span an unprecedented range of morphologies and physical properties, from  lenticulars galaxies (S0) to blue compact dwarfs (BCDs),  rotation velocities $(\sim 20 \kms{}\, \rm to \, \sim 300 \kms{})$, luminosities $(\sim  10^{7} L_{\odot}\, \rm to \,\sim 10^{12} L_{\odot})$, and gas content  $0.01\leq(M_{\HI{}}/L_{[3.6]}) \leq 10$.\\
We use the data available in the SPARC galaxy catalog to homogenously calculate and benchmark the stability levels in nearby galaxies. We then compare the stability of the nearby galaxies with the cold rotation supported galaxies at $z\,=\,4.5$, observed by \cite{rizzo2020dynamically,rizzo2021dynamical}. Since the two-component stability criterion $Q_{RW}$ considers the self-gravity of both the stars and gas, it enables us to ascertain the dominant mass component driving the stability levels. 
Further, $Q_{RW}>1$ ensures stability against axisymmetric instabilities, but a higher value of $Q_{RW}>Q_{critical}$ is needed for stabilizing the disc against non-axisymmetric perturbations and gas dissipation. \cite{griv2012stability} find that  $Q_{critical}\approx  2$ is needed for stability against non-axisymmetric instabilities, and further \cite{elmegreen2011gravitational} estimate $Q_{critical} \approx 2 \,-\, 3$ for stability against gas dissipation. We examine if the nearby galaxies are critically stable by estimating the minimum value of stability parameter $(Q^{Min}_{RW})$. 
In order to estimate where the galaxy is susceptible to growth of instabilities, we compute the radius at  which the $Q_{RW}$, $Q_{Stars}$ and $Q_{Gas}$ attain their minimum values given by $\big(R/R_{D}\big)_{Min(Q_{RW})}$,  $\big(R/R_{D}\big)_{Min(Q_{Stars})}$ and  $\big(R/R_{D}\big)_{Min(Q_{Gas})}$ respectively, where $R_{D}$ is the scalelength corresponding to the exponential surface density of the stellar disc. We then bin the galaxies by their morphological type and measure the stability/instability levels by computing $Q_{RW}$ as a function of the morphological type.We will then discuss the role of the gas fraction and the dark matter potential in regulating the stability levels. Finally, we note that, with the advent of the James Webb Space Telescope (JWST) and Square Kilometer Array (SKA), it will be possible to study the evolution of gravitational instabilities as a function of both morphological type and redshift.

The paper is organized as follows: we discuss the data and methods in \S 2. We present the results in \S 3 and discuss the implication of our results in \S 4. We will finally summarize our results in  \S 5.

\section{Data \& Method}
In this work, we consider a sample of $175$ galaxies taken from the \emph{Spitzer} Photometry and Accurate Rotation
Curves (SPARC) database\footnote{http://astroweb.cwru.edu/SPARC/} \citep{lelli2016sparc}. The
sample spans a wide range of morphologies from spirals to lenticulars and blue compact dwarfs with stellar photometry in the $3.6$  $\rm \mu m$  band and high-quality \HI{}+ H$\alpha$ rotation curves. 
The near-infrared band traces the stellar mass distribution very well as the mass-to-light ratio $(\gamma^{Stars})$ in 3.6 $\rm \mu m$  
is fairly constant for galaxies with varying masses and morphologies \citep{mcgaugh2014color}. We adopt value of $\gamma^{Stars}$ equal to 0.5 following \cite{mcgaugh2014color}.

In order to compute the stability, we use the two-component stability criterion introduced by \cite{romeo2011effective} 
\begin{equation}
\frac{1}{Q_{RW}} = 
                \begin{array}{ll} \frac{W_{\sigma}}{T_{Stars}Q_{Stars}} + \frac{1}{T_{Gas}Q_{Gas}}  \hspace*{0.5cm} if \hspace*{0.5cm}  T_{Stars}Q_{Stars} > T_{Gas}Q_{Gas}\\  
         \frac{1}{T_{Stars}Q_{Stars}} +\frac{W_{\sigma}}{T_{Gas}Q_{Gas}}                            \hspace*{0.5cm} if \hspace*{0.5cm} T_{Stars}Q_{Stars} < T_{Gas}Q_{Gas}  
         \end{array}
\end{equation}
\noindent and the weight function $W_{\sigma}$ is given by
\begin{equation}
 W_{\sigma} =\frac{2\sigma_{Stars} \sigma_{Gas}}{\sigma_{Stars} ^{2} + \sigma_{Gas}^ {2}}.
\end{equation}
\noindent The thickness correction is defined as
\begin{equation}
 T \approx 0.8 + 0.7 \frac{\sigma_{z}}{\sigma_{R}}
\end{equation}
\noindent where $\sigma_{z}$ and $\sigma_{R}$ are the vertical and the radial velocity dispersion, $\sigma_{Stars}$ and $\sigma_{Gas}$ are the velocity dispersion of the stars and gas. $Q_{Stars}$ and $Q_{Gas}$ correspond to the Toomre Q of the stellar and the gaseous disc. Toomre Q parameter is defined as $Q_{i}= \frac{\kappa \sigma_{i}  }{\pi G \Sigma_{i}}$, where $\kappa$ is the epicyclic frequency, and $\Sigma_{i}$ and $\sigma_{i}$ are the surface density and the radial velocity dispersion of either stars or gas.

A value of $Q_{RW} >1$ indicates that the galaxy is stable against the growth of axisymmetric perturbations. 
The epicyclic frequency $\kappa$ at a radius R is defined as 
\begin{equation}
\kappa^2(R)= \bigg( R\frac{d\Omega^{2} (R)}{dR} + 4\Omega^{2} (R)  \bigg)
\end{equation}
where $\Omega$ is the angular frequency defined as $\Omega^{2} (R)=\frac{1}{R}\frac{d\Phi_{Total}}{dR}= \frac{V^{2}_{rot}}{R^{2}}$, $\Phi_{Total}$ is the total gravitational potential and $V_{Rot}$ the total rotation velocity. We use the observed rotation curve from the SPARC catalog to compute the value of $\kappa$. We use the value of the inclination-corrected stellar luminosity as a function of radius from the SPARC database and multiply it by the $\gamma^{Stars}$ to derive the stellar surface density profile $(\Sigma_{Stars})$. 

We compute the radial velocity dispersion of the stars following the methods detailed in \cite{leroy2008star,romeo2017drives} and \cite{villanueva2021edge}. The radial velocity dispersion is defined as
\begin{equation}
\sigma_{Stars}(R)= \frac{1}{0.6} \sqrt{ \frac{2 \pi G R_{D}\Sigma_{Stars}(R)}{7.3}.}
\end{equation}
\noindent In the above equation, $R_{D}$ is the disc scalelength and is obtained by fitting the surface brightness profile with an 
exponential function $\Sigma_{Stars}(R)= \Sigma_{0}e^{-R/R_{D}}$. 

We derive the gas surface density ($\Sigma_{Gas}$) using the value of the circular velocity of gas $(V_{Gas})$ given in the  SPARC database,
\citep{mestel1963galactic, mera1996disk, binney2011galactic}
\begin{equation}
\Sigma_{Gas}= \frac{V^{2}_{Gas}}{2 \pi G R}.
\end{equation}
\noindent The value of $V_{Gas}$ has been multiplied by $1.33$ to correct for the presence of Helium and other metals.

We use a constant value of gas dispersion equal to  $10\, \kms$, which is consistent with the observations of the gas dispersion in nearby galaxies. 
\cite{mogotsi2016hi} measure mean \HI{} dispersion equal to $11.7 \pm 2.3\, \kms$ for a sample of nearby galaxies taken from high-resolution THINGS \HI{} survey \citep{walter2008things}, similarly \cite{tamburro2009driving} measure \HI{} dispersion equal to  $10 \pm 2 \,\kms{}$, also see \cite{romeo2020lenticulars}.

\noindent With all the ingredients needed for computing the two-component stability parameter ($Q_{RW}$) in place, we will present the results and analysis in the following section.

\section{Results}

\begin{figure*}
\resizebox{180mm}{70mm}{\includegraphics{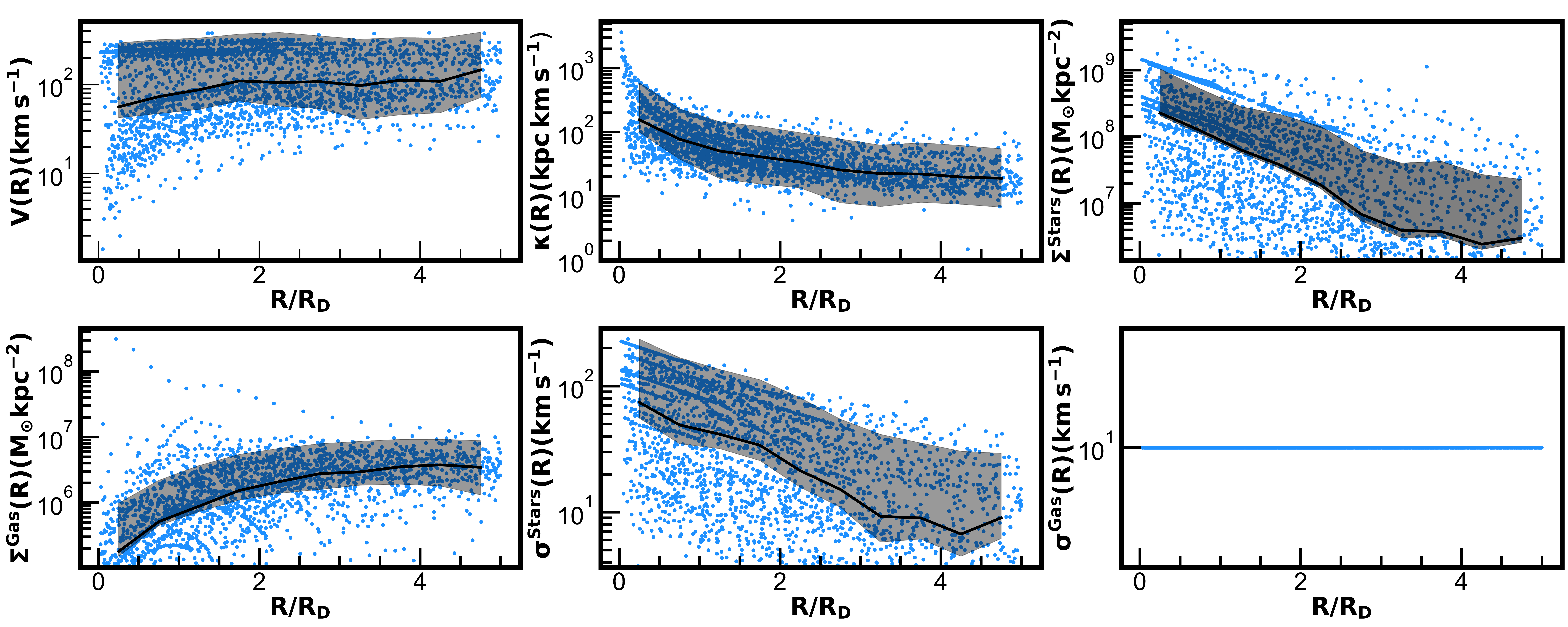}} 
\caption{Radial variation of the physical properties of the galaxies in the SPARC catalog. 
The solid line indicates the median and the shaded region indicates the $16^{th}$ and the $84^{th}$ percentile obtained by radially binning the data.}
\end{figure*}

In Figure 1, we show the properties of the galaxies in the SPARC sample as a function of the radius needed for computing the 
stability parameter ($Q_{RW}$). The median rotation velocity varies between  $56.2^{+224.4}_{-13.8} \leq \frac{V(R)}{\kms} \leq 146^{+280.4}_{-74.6}$,
i.e. the radially binned $16^{th}$ percentile of the rotation velocity lies between 13.8 \kms to 74.6 \kms whereas the 
$84^{th}$ percentile value of the rotation velocity varies between 224\kms to 280\kms, and the median rotation velocity varies between 56\kms to 146\kms.
The lower limits are the minimum and maximum of the $16^{th}$ percentile, and the upper limits are the minimum and the maximum of the 
$84^{th}$ percentile $\bigg( Min[Med.]^{+Min(84^{th})}_{-Min(16^{th})}\leq Quantity\leq Max[Med.]^{+Max(84^{th})}_{-Max(16^{th})}\bigg)$.
The median epicyclic frequency lies between  $18.9^{+35.8}_{-12.3} \leq \frac{\kappa(R)}{\rm kpc\kms} \leq 154.8^{+433.3}_{-54.8}$.
The radially binned median stellar surface density varies between 
$6.4^{+7.3}_{-5.6} \leq log_{10}\big(\frac{\Sigma_{Stars}}{M_{\odot}kpc^{-2}}\big)\leq 8.3^{+8.9}_{-7.3}$. The median gas surface density ranges between  $ 5.2^{+5.9}_{-4.4} \leq log_{10}\big(\frac{\Sigma_{Gas}}{M_{\odot}kpc^{-2}}\big)\leq 6.6^{+6.7}_{-6.3} $. 
The local median dispersion of the stars varies between $ 6.7^{+20.3}_{-2.2} \leq \frac{\sigma_{Stars}}{\kms}\leq 74.2^{+161.5}_{-18.6}$. We use a constant gas dispersion equal to 10\kms motivated by the observation of \HI{} dispersion in the nearby galaxies (see discussion in \S 2). The observed scatter in the input parameters reflects the diversity of the physical properties of the galaxies in the SPARC catalog.\\

\begin{figure*}
\resizebox{180mm}{38mm}{\includegraphics{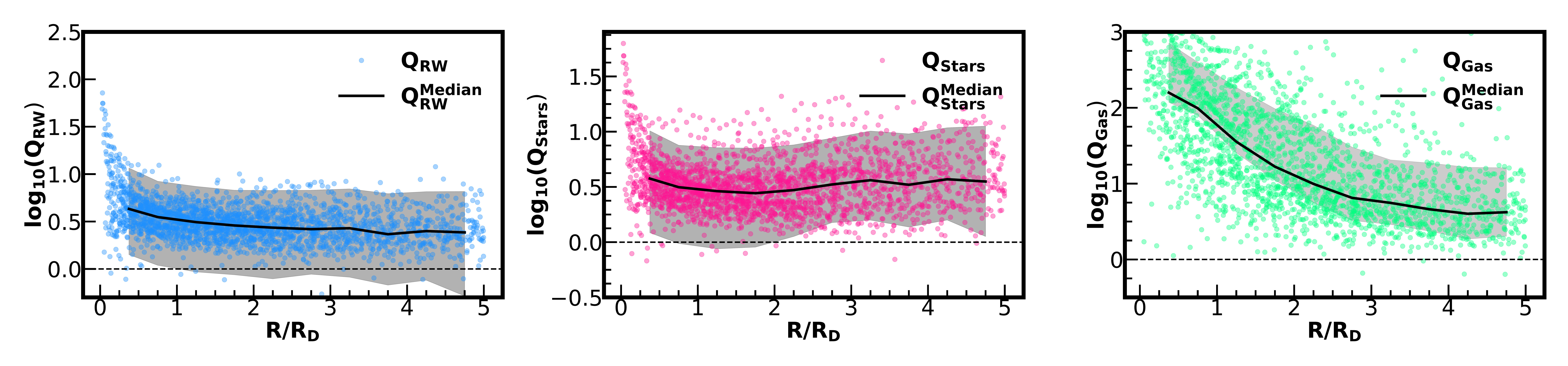}} 
\caption{We show the radial variation of the two-component stability parameter $(Q_{RW})$ in the first panel, and the stability of stellar and the gas 
components given by $Q_{Stars}$ and $Q_{Gas}$ respectively in the second and the third column.  The solid black line shows the median obtained by radially binning the data and the shaded region shows the median + $84^{th}$ percentile and the median - $16^{th}$ percentile in each radial bin.}
\end{figure*}

\subsection{Radial Variation of $Q_{RW}$}
In the first panel of Figure 2, we show the radial variation of the two-component stability parameter ($Q_{RW}$), followed by the stability parameter of stars $(Q_{Stars})$ and the gas $(Q_{Gas})$ in the second and the third panel respectively. The median value of $Q_{RW}$ varies between $2.32$ and $4.3$ $\bigg(2.32^{+3.4}_{-1.6}\leq Q^{Median}_{RW}\leq 4.3^{+7.3}_{-2.8}\bigg)$, whereas the median of $Q_{Stars}$ varies between $2.8$ and $3.8$ 
$\bigg(2.8^{+4.3}_{-1.8} \leq Q^{Median}_{Stars}\leq 3.8^{+6.4}_{-2.5}  \bigg)$. 
The local median of $Q_{Gas}$ varies in between $4.0$ and $158.2$ 
$\bigg(4.0^{+12.0}_{-2.1} \leq Q^{Median}_{Gas}\leq 158.2^{+552}_{-40}\bigg)$. 
The median value of $Q_{Gas}$ is high at the center and is comparable to the value of $Q_{Stars}$ at the outer radius ($>3.5R_{D}$). The high stability of the gas disc at the center is due to a large value of the epicyclic frequency and a small value of the gas surface density.
The median value of $\Sigma_{Gas}$ at the center is only $0.08\%$ relative to the median value of $\Sigma_{Star}$ at the center. So, the contribution of the gas disc to the total self-gravity at the center is negligible compared to the contribution of the stellar disc. Hence, $Q_{RW}$ in the inner region up to $2R_{D}$ follows $Q_{Stars}$, see Figure 3.  In other words, the net stability in the inner region is driven by the stellar disc. The gas dispersion  at 2$R_{D}$ is  ($\sigma_{Gas}=10 \kms$) comparable to the value of the stellar dispersion ($\sigma_{Stars}= 9.2 \kms$), but a small value of $\sigma_{Gas}$ is insufficient to effect the total stability.  In the following section (\S 3.2), we test the effect of adopting a lower value of gas dispersion motivated by the observed velocity dispersion of the molecular gas.  \\
Beyond $2R_{D}$, the median stability of stars $\bigg(Q^{Med}_{Stars}(2R_{D})=2.9\bigg)$ becomes more than the two-component stability parameter $\bigg(Q^{Med}_{RW}(2R_{D}) =2.75\bigg)$ as the surface density of the stellar disc falls off faster than the epicyclic frequency($\kappa$) and radial velocity dispersion ($\sigma_{Stars}$). The median epicyclic frequency $(\kappa)$ and the stellar dispersion $(\sigma_{Stars})$ are $14.5\%$ and $12.4\%$ at $2R_{D}$ relative to their corresponding values at the center. However, the value of stellar surface density at $2R_{D}$ is only 1.7$\%$ relative to its value at the center. \\
Finally, we find that the median values of $Q_{Gas}$ are consistent with the critical surface density of the gas disc. \cite{lin1993protostars,wang1994gravitational,boissier2003star, burkert2013dependence} show that the star formation takes place above a gas surface density called critical gas surface density.
The critical gas density is defined as \citep{wang1994gravitational} 
\begin{equation}
\Sigma_{critial}=\gamma \frac{\kappa \sigma_{Gas}}{\pi G} \, \rm and \,   \gamma=\bigg(1+ \frac{\Sigma_{Stars}\sigma_{Gas} }{\Sigma_{Gas}\sigma_{Stars} }   \bigg)^{-1}. 
\end{equation}
The critical gas density can be further written as $\Sigma_{Gas}/\Sigma_{Critical}=1/(\gamma Q_{Gas})$. The median value of $\gamma$ for the galaxies in the SPARC sample is equal to $0.25$. The median value of $Q^{Min}_{Gas}$ is equal to $2.45$, which gives a value of $\Sigma_{Gas}/\Sigma_{Critical}\approx 0.6$, consistent with the values obtained by \cite{boissier2003star} for a sample of 16 spiral galaxies. Thus, a higher value of $\Sigma_{Gas}/\Sigma_{Critical}$ driven by a smaller value of  $Q_{Gas}$ will induce favorable conditions for the onset of gravitational instabilities.

\subsection{Effect of $\sigma_{Gas}$ on $Q_{RW}$}
In Figure 3, we see that the $Q_{RW}$ in the inner radii is strongly regulated by the stellar component, whereas in the outer region, the effect of $Q_{Gas}$ on $Q_{RW}$ becomes important. This effect arises primarily due to negligible gas surface densities in the inner region compared to the stellar surface density. Since the epicyclic frequency, $\kappa$ has a similar effect on both the stellar and the gas disc; it becomes imperative to check the effect of the gas dispersion on the radial variation of $Q_{RW}$. We show the effect of adopting a smaller value of gas dispersion equal to 6\kms on the median stability of the galaxies in the SPARC catalog in Figure 3. The value of the gas dispersion equal to 6\kms is consistent with the velocity dispersion of the molecular gas measured in the Milky Way $\sigma =4.4 \pm 1.2$ \kms \citep{marasco2017distribution}. From, Figure 3, it is evident that a smaller gas dispersion lowers the value of $Q_{Gas}$; see the green dashed line in Figure 3. But, the $Q_{RW}$ in the inner region is still primarily driven by the stellar component despite a small value of the gas dispersion. However, a small value of $Q_{Gas}$ makes the value of $Q_{RW}$ smaller in the outer region of the galaxies when $\sigma_{Gas}=6\kms$. Thus, a cold gas disc aided by a high gas surface density can potentially decrease the value of $Q_{RW}$, making the outer region of nearby galaxies more susceptible to the growth of axisymmetric instabilities. Overall we find that the global median of $Q_{RW}$ is unaffected by a smaller value of gas dispersion. $Q_{RW}$ changes from 3.0 ($\sigma_{Gas}=10\kms$) to 2.8 upon adopting a lower gas dispersion ($\sigma_{Gas}=6\kms$).

\begin{figure}
\resizebox{80mm}{60mm}{\includegraphics{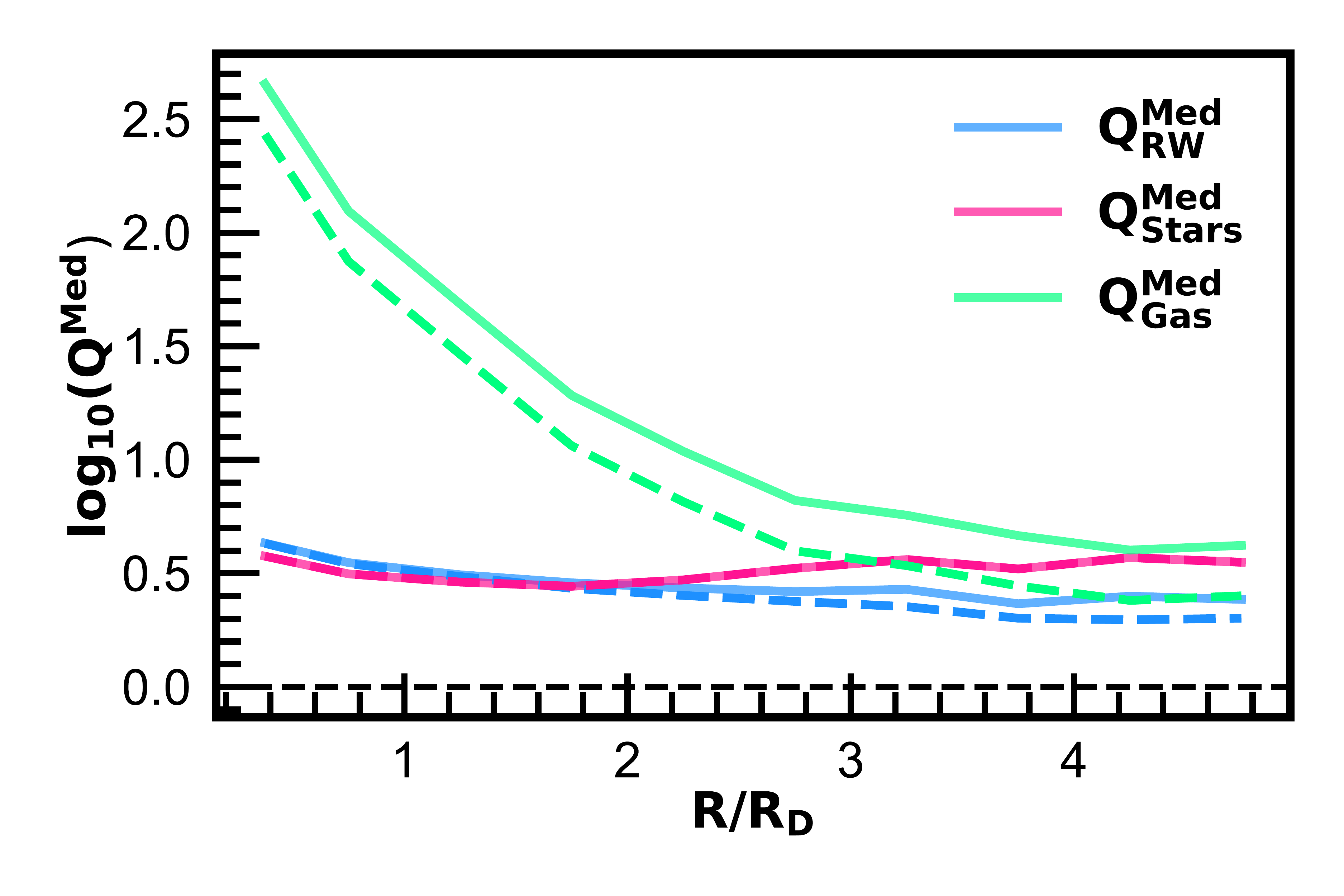}} 
\caption{The solid lines depicts the radial variation of the median stability parameter when the gas dispersion is equal to 10\kms, whereas the dashed line depicts the radial variation of the median stability parameter when the gas dispersion is equal to 6\kms. The black dashed line shows the value of marginal stability $Q_{RW}=1$.}
\end{figure}

\subsection{Critical Stability}
A value of $Q_{RW}>1$ indicates that the galaxy is stable against axisymmetric instabilities. However, higher value of $Q_{RW}>Q_{critical}$ is needed 
to stabilize the galaxy against non-axisymmetric instabilities \citep{griv2012stability} and gas dissipation \citep{elmegreen2011gravitational}. 
\cite{griv2012stability} and \cite{elmegreen2011gravitational} estimate the value of $Q_{critical}$ to be between 2 - 3 for the galaxy to be stable against non-axisymmetric instabilities and gas dissipation. In order to understand if the galaxies in the local universe are critically stable, we inspect the minimum values of total stability given by $Q^{Min}_{RW}$, and the minimum values of the stability parameter for stars and gas given by $Q^{Min}_{Stars}$ and $Q^{Min}_{Gas}$ respectively. A value of $Q^{Min}_{RW}$ in the range of $2\,-\,3$ indicates that the galactic disc is critically stable, indicating stability against gas dissipation and non-axisymmetric instabilities as pointed out by \cite{griv2012stability} and \cite{elmegreen2011gravitational}. 
The median value of $Q^{Min}_{RW}$ for the nearby galaxies is equal to 2.2, whereas the median value of  $Q^{Min}_{Stars}$ and $Q^{Min}_{Gas}$ is equal to $2.4$ and $2.45$ respectively. We note that $91\%$ of galaxies in the SPARC catalog have $Q^{Min}_{RW} >1$, $94\%$ of galaxies have $Q^{Min}_{Stars} >1$ and $Q^{Min}_{Gas} >1$. This indicates that 91$\%$ galaxies in the SPARC catalog are stable against axisymmetric instabilities at all radii. Further, $56\%$ of the galaxies in the sample have $Q^{Min}_{RW}$ in the range of $2 - 3$, which ensures critically stability at all radii. The SPARC sample consists of 81 irregular $(Type= 8 - 11)$ and 94 spiral galaxies $(Type= 0 - 7)$. We note that 63 out of 81 irregular galaxies have a $Q^{Min}_{RW} >2$ and only 36 spiral galaxies out of 94 have a $Q^{Min}_{RW} >2$. Further, 15 galaxies in the SPARC catalog have $Q^{Min}_{RW} <1$, of which 5 are irregular galaxies and 10 are spiral galaxies.
We note that a large number of irregular galaxies in SPARC catalog are stable against gas dissipation and non-axisymmetric instabilities.

\subsection{What about galaxies with $Q^{Min}_{RW}<1$ ?}
\begin{figure}
\resizebox{85mm}{60mm}{\includegraphics{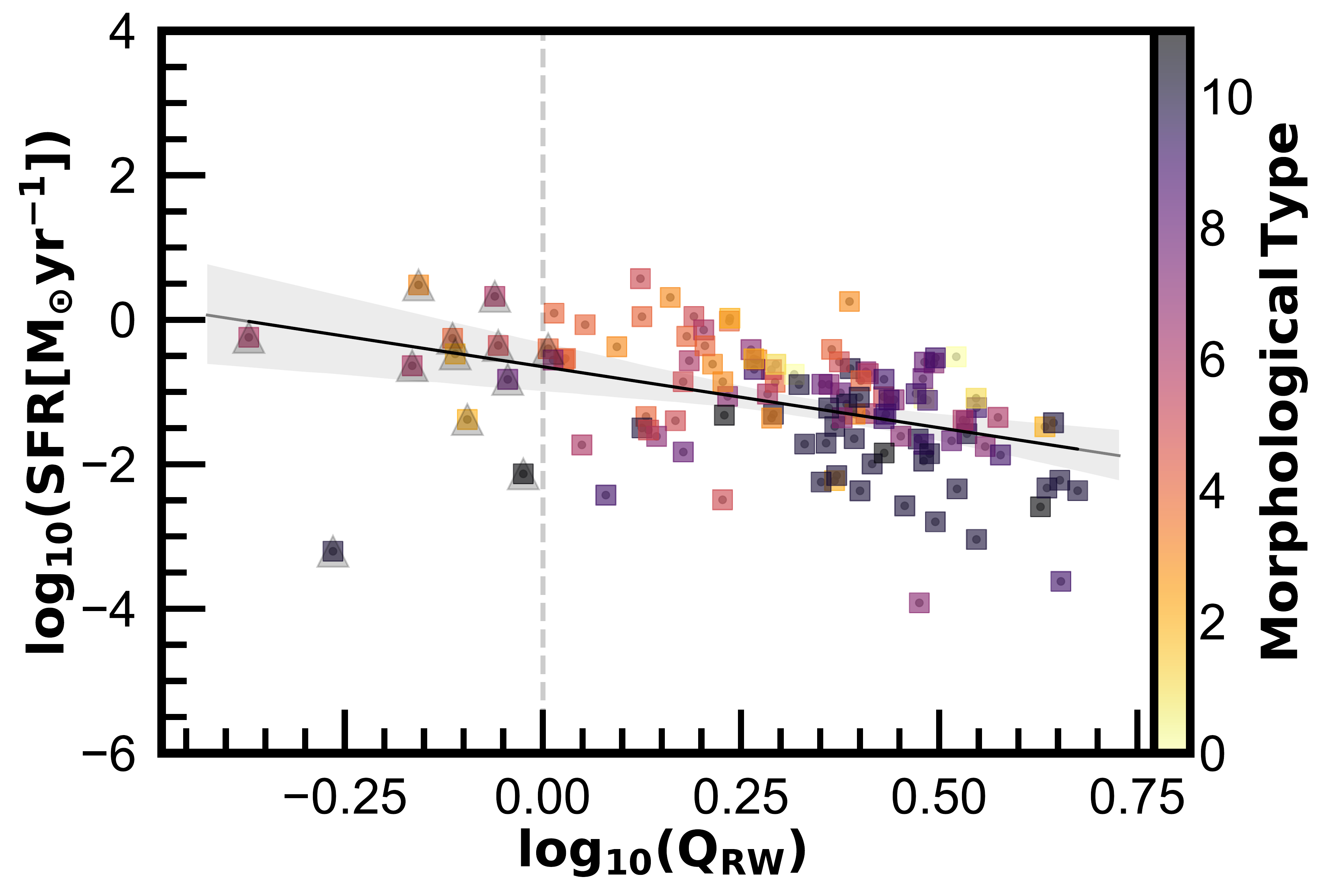}} 
\caption{$Q^{Min}_{RW}$ versus the star formation rate for a sample of $120$ galaxies for which NUV magnitudes were available in GALEX. The galaxies are color-coded according to their morphological type [Type 0=S0, 1=Sa, 2 = Sab, 3 = Sb, 4 = Sbc, 5 = Sc, 6 = Scd, 7 = Sd, 8 = Sdm, 9 = Sm, 10 = Im, 11 = BCD]. Types = 0 - 7 are spiral galaxies and Type = 8 - 11 are irregular galaxies.
The triangles depict the galaxies which have a $Q^{Min}_{RW}<1$. The vertical $'dashed'$ line shows $Q_{RW}=1$. The $'black'$ line depicts the regression line, and the shaded region indicates the 95$\%$ confidence interval. }
\end{figure}

A direct consequence of $Q_{RW}<1$ is that the galactic disc becomes unstable against the growth of axisymmetric instabilities. In order to investigate the effect of the $Q^{Min}_{RW}<1$ for galaxies in our sample, we study the star formation rate, which is a direct outcome of gravitational instability. We query the near ultra-violet (NUV) magnitudes 
from Galaxy Evolution Explorer (GALEX) \footnote{https://galex.stsci.edu/} database using the astroquery\footnote{https://www.astropy.org/astroquery/} service. We find the NUV magnitudes for $120$ galaxies out of the $175$ galaxies in the SPARC catalog and convert the GALEX magnitudes to the star formation rate following the prescription given by \cite{kennicutt1998star}
\begin{equation}
SFR(M_{\odot}\, yr^{-1})= 1.4 \times 10^{-28} L_{NUV} (ergs\, \,s^{-1}\,Hz^{-1}).
\end{equation}
In Figure 4, we show the star formation rate calculated using the NUV magnitudes for $120$ nearby galaxies versus the $Q^{Min}_{RW}$. The plot shows a negative relationship between 
$Q^{Min}_{RW}$ and the star formation rate. The points marked with $'triangles'$ represent galaxies that have 
$Q^{Min}_{RW}<1$. The median value of star formation rates (SFR) for galaxies with $Q^{RW}_{Min}<1$ is equal to $0.4 \, M_{\odot} yr^{-1}$, whereas SFR for galaxies with $Q^{Min}_{RW}>1$ is equal to $0.07 \, M_{\odot} yr^{-1}$. As expected, the galaxies with $Q^{Min}_{RW}< 1$ have higher star formation rate than those with $Q^{Min}_{RW} > 1$. Further, we report a negative relation between the star formation rates and the minimum value of the stability parameter given by 

\begin{equation}
log(SFR/M_{\odot}yr^{-1})=-0.65 - 1.69 log(Q^{Min}_{RW}).
\end{equation}

We establish an anti-correlation between the global star formation rate and the stability parameter. Our results are consistent with the results previously obtained by \cite{westfall2014diskmass}, who show that the star formation surface densities anti-correlate with the value of $Q_{RW}$ at $1.5R_{D}$ for a sample of $27$ galaxies.
Further, we can see from Figure 4 that spiral galaxies (Type=0 - 7) and irregular galaxies (Type= 8 - 11) arrange themselves in two distinct clusters. Typically the spiral galaxies have a higher value of median star formation rate $(SFR=0.2 M_{\odot}yr^{-1})$ compared to the irregular galaxies, which have a median star formation rate equal to 0.02 $M_{\odot}yr^{-1}$. The observed star formation rate in the spirals and irregular galaxies is consistent with the measured values of the $Q_{RW}$. The spiral galaxies typically have lower median stability ($Q^{Min}_{RW}=1.7$) and hence a higher star formation rate compared to the irregular galaxies ($Q^{Min}_{RW}=2.6$).

\subsection{Where do galaxies become unstable ?}
In order to answer the above question, we estimate the radius at which $Q_{RW}$, $Q_{Stars}$, and $Q_{Gas}$ attain their minimum value, 
given by $\big(R/R_{D}\big)_{Min(Q_{RW})}$, $\big(R/R_{D}\big)_{Min(Q_{Stars})}$ and $\big(R/R_{D}\big)_{Min(Q_{Gas})}$ respectively.
The median value of the $\big(R/R_{D}\big)_{Min(Q_{RW})}$ is equal to  $2.8$, and the median values of  
$\big(R/R_{D}\big)_{Min(Q_{Stars})}$ and $\big(R/R_{D}\big)_{Min(Q_{Gas})}$ are equal to  $1.4$ and $5.1$ respectively.
The stellar disc has minimum stability close to $1.4$ times the disc scalelength, i.e. at a point where the stellar density becomes $\rm e^{-1.4}$ of its value at the center. On the other hand, the gas disc is much more extended than the stellar disc and has a minimum stability at $5.1$ times the disc scalelength of stars. 
However, The composite stability parameter taking into account the self-gravity of both the stars and gas attains a minimum value at $2.8$ times the disc scalelength of the stars. This highlights the importance of taking into account the self-gravity of stars and gas consistently. If one were to consider the stability of just the stellar disc, the galaxies would have minimum  stability at $1.4R_{D}$  close to the center, 
on the other hand, if one were to consider just the gas disc, the galaxies will have minimum stability at $5.1R_{D}$, away from the center. However, the composite stability parameter indicates that the galaxies attain minimum stability at an intermediate value equal to $2.8R_{D}$.

\subsection{Stability versus morphological type}

\begin{figure*}
\hspace*{-2cm}
\resizebox{220mm}{95mm}{\includegraphics{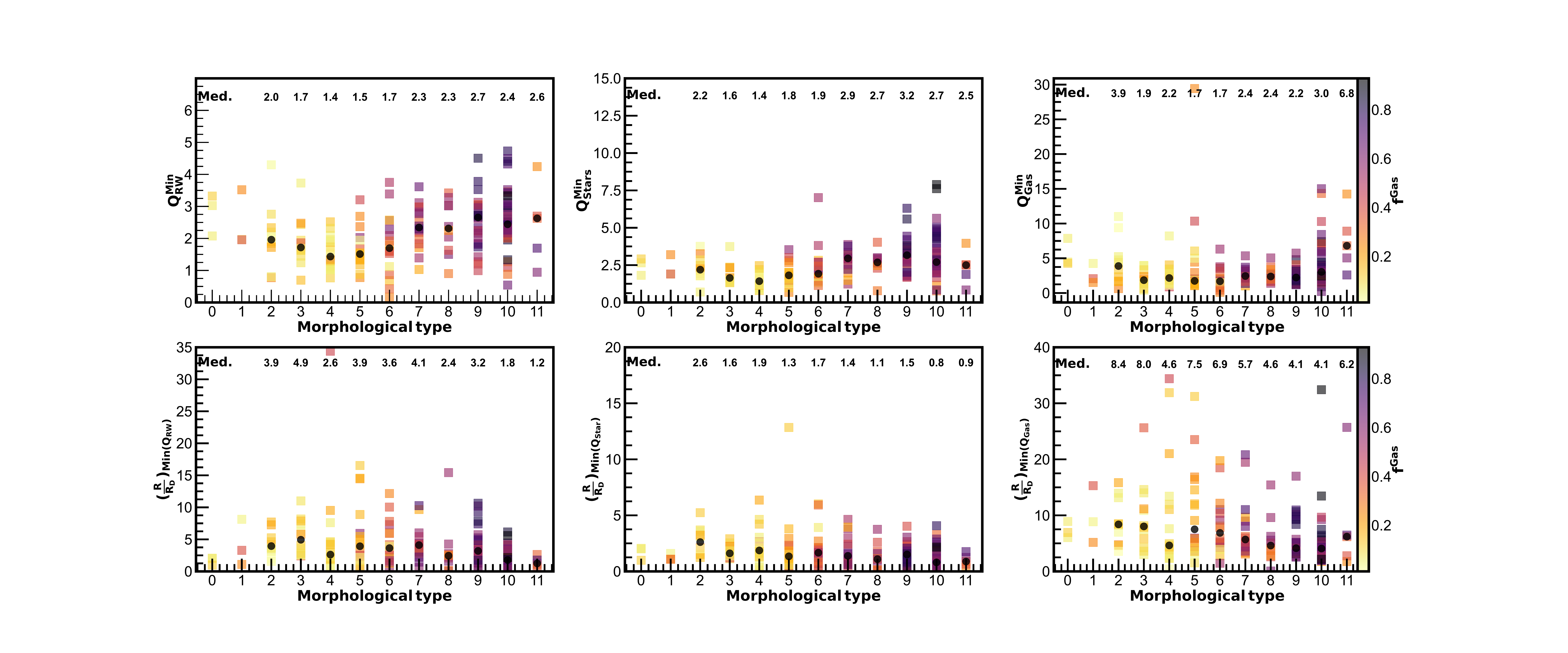}} 
\caption{In the top panel we show $Q^{Min}_{RW}$, $Q^{Min}_{Stars}$ and $Q^{Min}_{Gas}$ 
as a function of the morphological type. In the lower panel, we show the radius at which the $Q_{RW}$, $Q_{Stars}$ and $Q_{Gas}$ attain their minimum value as a function of the morphology. The $'black'$ marker indicates the median value of the quantity. We use the following convention for the galaxy classification Type 0=S0, 1=Sa, 2 = Sab, 3 = Sb, 4 = Sbc, 5 = Sc, 6 = Scd, 7 = Sd, 8 = Sdm, 9 = Sm, 10 = Im, 11 = BCD. The galaxies are color coded according to the gas fraction $(f^{Gas})$.}
\end{figure*}

In order to understand how stability varies from one morphological type to another, we bin the galaxies according to their morphological type and compute $Q^{Min}_{RW}$, $Q^{Min}_{Stars}$, 
and $Q^{Min}_{Gas}$. Further, we measure the radius at which the galaxies attain a minimum value of $Q_{RW}$, $Q_{Stars}$, and $Q_{Gas}$. We present the results in Figure 5. In Figure 5, we adopt the convention for galaxy classification following \cite{lelli2016sparc}; Type 0=S0, 1=Sa, 2 = Sab, 3 = Sb, 4 = Sbc, 5 = Sc, 6 = Scd, 7 = Sd, 8 = Sdm, 9 = Sm, 10 = Im, 11 = BCD. Galaxy types numbered 1 - 7 are spirals, and 8 - 11 are irregular galaxies. In the top panel of Figure 5, we show the median values of  $Q^{Min}_{RW}$, $Q^{Min}_{Stars}$, and $Q^{Min}_{Gas}$, for galaxies in each morphological bin. In the bottom panel, we show the median value of $R/R_{D}$ at which $Q_{RW}$, $Q_{Stars}$ and $Q_{Gas}$ attain their minimum value as a function of the morphology. We leave out the galaxies with Type= 0 and Type= 1 as there are only 2 galaxies of Type=0 and 3 galaxies of Type=1. \\

\textbf{\underline{$Q_{RW}$ versus morphological type}}\\
In panel 1 of Figure 5, we show the $Q^{Min}_{RW}$ for galaxies of different morphological types in the SPARC catalog. The median value of $Q^{Min}_{RW}$ 
varies between $1.4$ for Type 4 (Sbc) and $2.7$ for Type 9 (Sm), indicating that both spirals and irregulars are stable against axisymmetric instabilities. 
As we move across the Hubble sequence; Sab (Type=2) galaxies have $Q^{Min}_{RW}=2.0$ and BCDs (Type=11), have $Q^{Min}_{RW}=2.6$. Further, we find that Sab (Type=2) galaxies attain minimum $Q_{RW}$ at $3.9R_{D}$, whereas BCDs (Type=11) and Im(Type=10) attain minimum $Q_{RW}$ at $1.2R_{D}$ and $1.8R_{D}$ respectively. The median radius at which the $Q_{RW}$ attains a minimum value ranges between $1.2$ for BCDs (Type=11) and $4.9$ for Sb (Type=3) galaxies. 

Although both the spiral and the irregular galaxies in the sample are stable against the growth of axisymmetric instabilities ($Q^{Min}_{RW}>1$), we find that the spiral galaxies are characterized by smaller values of $Q_{RW}=1.7$ and a median gas fraction $(f^{Gas}=0.1)$  compared to the irregular galaxies which have a median $Q_{RW}=2.6$ and $f^{Gas}=0.6$ . The small value of $Q^{Min}_{RW}$ and $f^{Gas}$ observed in the spiral galaxies is consistent with the star formation rates observed in these galaxies. A small value of $f^{Gas}$ in spiral galaxies indicates that the gravitational instabilities efficiently convert the available gas into stars. \\

\textbf{\underline{$Q_{Stars}$ versus morphological type}}\\
The median $Q^{Min}_{Stars}$ varies between $1.4$ for Sbc (Type =4) galaxies and $3.2$ for Sm (type =9) galaxies. The median value of $Q^{Min}_{Stars}$ for Sab (Type=2) is equal to $2.2$, and for
BCDs (Type=11) is equal to $2.5$. The stellar disc, in the case of Im and BCDs attains $Q^{Min}_{Stars}$ at 0.8$R_{D}$ and 0.9$R_{D}$ respectively. The Im (Type = 10) galaxies become unstable at $0.8R_{D}$ whereas Sa (Type =2) become unstable at $2.6R_{D}$. 
We note that the median $Q^{Min}_{Stars}$ for the spiral galaxies $(Q^{Min}_{Star})=1.9)$ is higher than the $Q^{Min}_{Stars}$ obtained for the irregular galaxies $(Q^{Min}_{Star}=2.8)$.\\

\textbf{\underline{$Q_{Gas}$ versus morphological type}}\\
The $Q^{Min}_{Gas}$ is equal to 1.7 for Sc(Type =5) and Scd (Type =6) galaxies and is equal to 6.8 for BCDs (Type =6). The Sm and Im galaxies attain $Q^{Min}_{Gas}$ at $4.1R_{D}$ and the Sab galaxies at $8.4R_{D}$.
Further, $Q_{Gas}$ attains a minimum value outside $3R_{D}$, beyond which the contribution of the stellar surface density becomes negligible. \\
Also, we note that across morphological types from Sa to BCDs $Q_{Stars}$ attains minimum between $0.8R_{D}$ to $2.6R_{D}$, whereas the $Q_{Gas}$ attains a minimum value outside $3R_{D}$. If one considers just the stellar disc and neglects the gas disc, the stellar disc attains minimum values close to the center. 
On the other hand, if we consider just the gas disc neglecting the contribution of the stellar disc, we find that the gas disc becomes unstable away from the center. However, when we use two-component stability parameter $(Q_{RW})$ taking into account the contribution of both the stellar and the gas disc on equal footing, we find that $Q_{Gas}$ drives the value of $Q_{RW}$ outwards.\\

We find that the irregular galaxies are more stable than the spiral galaxies, consistent with the higher gas fractions observed in the irregular galaxies than the spirals. This suggests that the gravitational instabilities are inefficient at converting the gas into stars in irregular galaxies consistent with the observed star formation rates $(SFR =0.02 M_{\odot}yr^{-1})$. On the other hand, the spiral galaxies have a lower median $Q^{Min}_{RW}$, a lower gas fraction, and a higher star formation rate $(SFR = 0.2 M_{\odot}yr^{-1} )$ indicating that the gravitational instabilities efficiently convert gas into stars. Finally, we note that the irregular galaxies attain $Q^{Min}_{RW}$ at $2.4R_{D}$, closer to the center, and the spiral galaxies attain minimum stability at $(3.6R_{D})$ from the center.

\section{Discussion \& Conclusions}
\begin{figure*}
\hspace*{-1.2cm}
\resizebox{200mm}{80mm}{\includegraphics{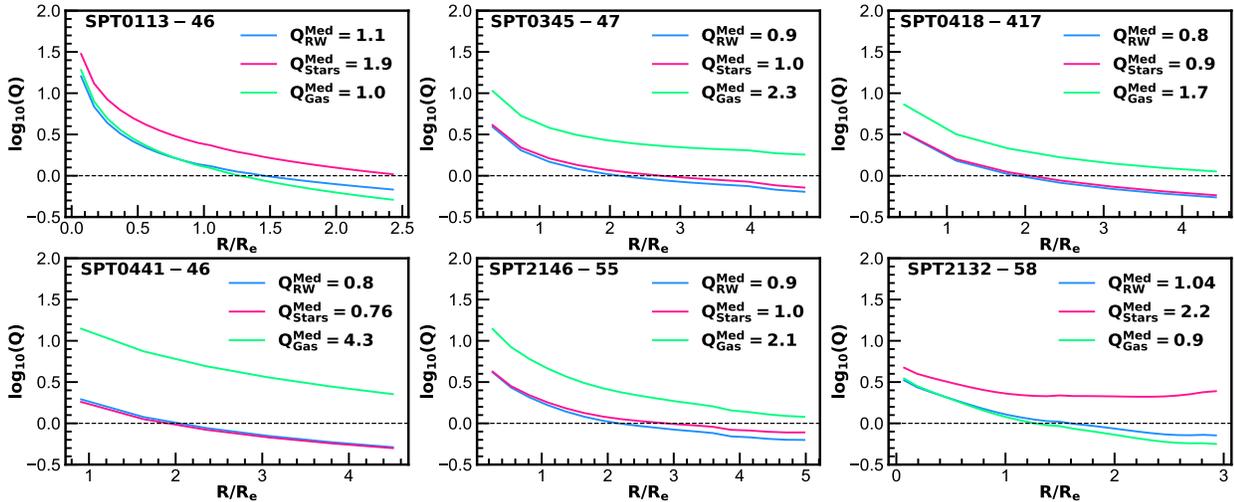}} 
\caption{The radial variation of the stability parameter $Q_{RW}$, $Q_{Stars}$ and $Q_{Gas}$ for the galaxies at redshift $z\,=\,4.5$. Further, we have shown the global median values of $Q_{RW}$, $Q_{Stars}$ and $Q_{Gas}$. The dashed line indicates the marginal stability $Q_{RW}=1$.}
\end{figure*}

\begin{figure}
\hspace*{-4mm}
\resizebox{80mm}{58mm}{\includegraphics{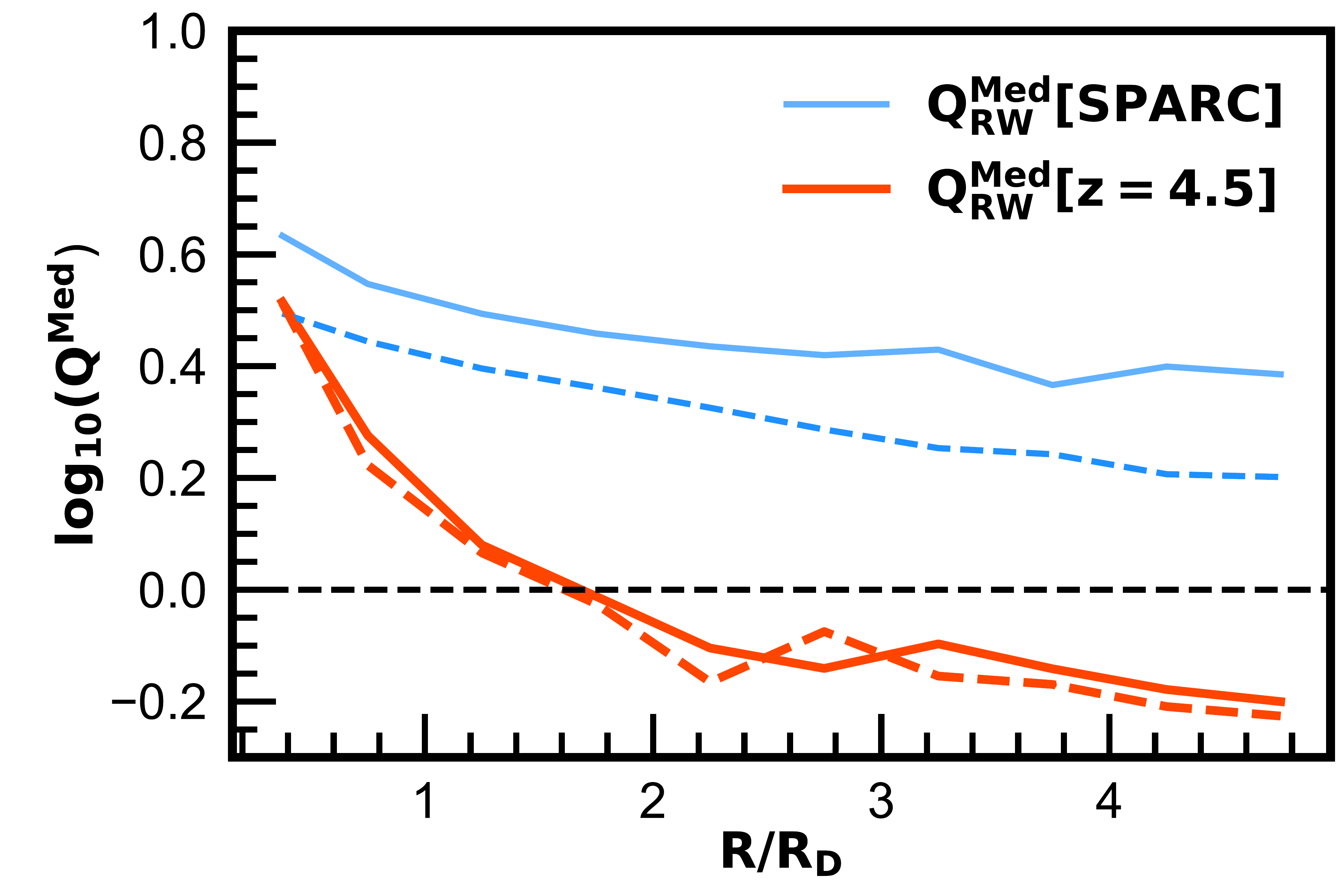}} 
\caption{The $'blue'$ and the $'red'$ lines indicate the radially binned median $Q_{RW}$ for galaxies from the SPARC catalog and the galaxies at $z\,=\,4.5$ respectively. The dashed line indicates the median $Q_{RW}$ upon removing the dark matter's contribution to the total potential.}
\end{figure}

\textbf{\underline{Stability of high redshift galaxies}}\\
The galaxies observed at high redshift are precursors to the galaxies in the local universe. Comparing the stability of the galaxies in the local universe to those observed at high redshift provides important information about how gravitational instabilities evolve as a function of redshift.
We compare $Q_{RW}$ obtained for local galaxies,  with a sample of $6$ galaxies observed at a 
redshift $z\,=\,4.5$ taken from \cite{rizzo2020dynamically,rizzo2021dynamical}. The value of 
rotation to random motion ($V/\sigma$) for the galaxies at $z\,=\,4.5$ lies between $7\,-\,15$, indicating that galaxies in the early universe are rotation supported and dynamically cold. 
We take the structural and kinematic properties of the galaxies at $z\,=\,4.5$ from \cite{rizzo2020dynamically,rizzo2021dynamical} and compute their stability following the methods detailed in \S 2. From, Figure 6, we see that the galaxies at $z\,=\,4.5$ are closer to the marginal stability ($Q_{RW}=1$)  with a median $Q_{RW}$ equal to 0.98. We note that the median $Q_{RW}$ at $z=4.5$ is lower than $Q_{RW}$ of the nearby galaxies. The low stability of the galaxies at $z\,=\,4.5$ is consistent with the high star formation rate of the order $10^{2} - 10^{3}\, M_{\odot}yr^{-1}$ \citep{rizzo2020dynamically,rizzo2021dynamical}. The SFR in galaxies at $z=4.5$ is significantly higher than the nearby galaxies in SPARC catalog, which have a median star formation rate equal to $0.07M_{\odot}yr^{-1}$. 
However, we note that nearby galaxies in the SPARC catalog, which have $Q_{RW}\leq 1$, have median star formation rate equal to $0.4\, M_{\odot}yr^{-1}$. 
Further, we find that in the case of SPT0113-46 and SPT2132-58, the $Q_{RW}$ values closely follow $Q_{Gas}$, whereas $Q_{RW}$ for all the other galaxies at $z\,=\,4.5$ is driven by $Q_{Stars}$. In other words, even in galaxies observed at $z=4.5$, the stellar disc drives the net stability similar to the trend observed in local galaxies. The consonance between the $Q_{RW}$ and $Q_{Stars}$ indicates that once the stellar disc has formed the stability levels are driven primarily by the stars suggesting the presence of an inherent mechanism that self-regulates the stability. 

Further, we inspect the value of gas fraction in the nearby spiral galaxies with a $Q_{RW}<1$. We find that the spiral galaxies (Type = 1 - 7), which have $Q^{Min}_{RW}<1$, have a gas fraction between $0.06$ and $0.34$. Whereas the galaxies at $z=4.5$ have a median gas fraction equal to 0.5
and $Q_{RW}<1$. Thus, a large gas fraction, a small median value of $Q_{RW}$, and a high star formation rate suggest that the galaxies observed at $z=4.5$ are currently undergoing star formation. The nearby spiral galaxies have relatively higher $Q_{RW}$, and a lower gas fraction and a star formation rate lower than galaxies at $z=4.5$, suggesting that the nearby spirals have reached a threshold star formation rate. \\

\textbf{\underline{Role of the dark matter on stability levels}}\\
In order to gain insight into the self-regulation mechanism, we derive the $Q_{RW}$ of the $'star+gas'$ disc by neglecting the contribution of the dark matter to the total potential. The effect of the dark matter on $Q_{RW}$ is encapsulated in the epicyclic frequency $(\kappa)$. 
The total epicyclic frequency can be written as $\kappa^{2}_{Total}= \kappa^{2}_{Stars} +\kappa^{2}_{Gas} +\kappa^{2}_{Dark\, Matter}$, where $\kappa_{Total}$ is derived from the observed rotation curve. The total epicyclic frequency $(\kappa_{Total})$  can be derived by adding in quadrature the contribution from the stars, gas, and the dark matter halo, respectively. 
This is possible as $\kappa$ is defined in terms of the angular frequency $(\Omega)$, which in turn is defined in terms of the potential, and the total potential can be written as, $\Phi_{Total}= \Phi_{Stars} + \Phi_{Gas} + \Phi_{Dark\, matter}$ (See Equation 5, and the paragraph beneath). We derive $Q_{RW}$ for the sample of nearby galaxies and galaxies at $z\,=\,4.5$ by neglecting the contribution of the dark matter i.e. $\kappa^{2}_{Total}= \kappa^{2}_{Gas} +\kappa^{2}_{Stars}$. 
From, Figure 7, we can see that the dark matter has a negligible effect on the median stability of the galaxies at $z \,= \,4.5$ (see the red lines). The global median of $Q_{RW}$, in the presence of dark matter, is equal to $0.99$, and upon removing the contribution of the dark matter from the total potential, the median $Q_{RW}$ changes to $0.93$. \cite{genzel2017strongly} in their observational study of high redshift galaxies $(z=0.6 -2.6)$ find that the massive disc galaxies at high redshift are dominated by baryons with a negligible dark matter fraction.
On the other hand, in the case of nearby galaxies, the overall stability decreases upon eliminating the dark matter's contribution to the total potential. The median value of $Q_{RW}$ decreases from $3.0$ in the presence of dark matter to $2.3$ in the absence of dark matter. Although the overall stability curve is lowered upon eliminating the dark matter's contribution to the total potential, the nearby galaxies continue to be stable against the growth of axisymmetric instabilities. This indicates that the $'star+ gas'$  disc can self-regulate the stability levels, atleast in a statistical sense.

\textbf{\underline{What about Low Surface brightness galaxies ? }}\\
\cite{garg2017origin} showed that the low surface brightness galaxies become susceptible to the growth of gravitational instabilities $(Q^{Min}_{RW}=0.7 - 1.5)$ upon removing the contribution of the dark matter. We 
find that LSBs in the SPARC catalog have a median stability equal to 2.4, and upon removing the contribution of the dark matter to the total potential, the $Q^{Min}_{RW}$ becomes 1.5, consistent with the previous study by \cite{garg2017origin}, also see \cite{narayanan2022superthin}.

\begin{figure*}
\resizebox{185mm}{47mm}{\includegraphics{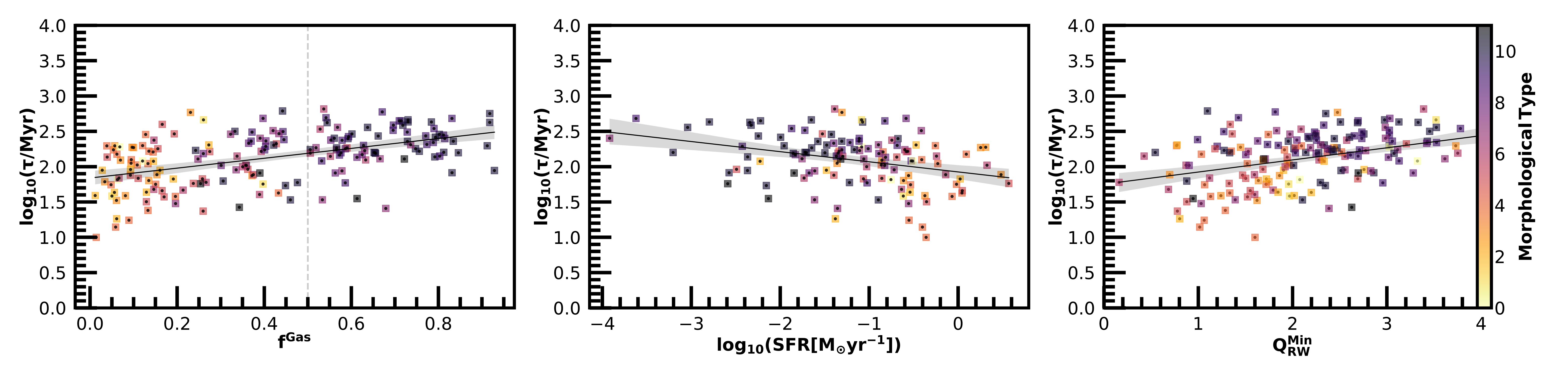}}
\caption{ In the first panel, we plot the time scale for the growth of gravitational instabilities $(\tau)$ as a function of $f^{Gas}$. In the second panel, we show $\tau$ as a function of the star formation rate (SFR) and plot $\tau$ as a function of $Q^{Min}_{RW}$ in the third panel. The vertical dashed line indicates $f^{Gas} =0.5$.}
\end{figure*}

\textbf{\underline{Connection between SFR, $f^{Gas}$ and $Q_{RW}$ }}\\
In this study, we find that the galaxies which have a smaller value of $Q_{RW}$ have a higher star formation rate and relatively a lower gas fraction. The spiral galaxies in our sample have a lower median stability $(Q^{Min}_{RW}=1.7)$ compared to irregular galaxies $(Q^{Min}_{RW}=2.6)$. Further, the stability levels are consistent with the  star formation rates in spiral galaxies $(SFR =0.2 M_{\odot}yr^{-1})$ and irregular galaxies $(SFR =0.02 M_{\odot}yr^{-1})$. We then inspect the gas fraction in the nearby galaxies and find that the spiral galaxies have a median $f^{Gas}$ equal to 0.1, and the irregular galaxies have a median $f^{Gas} =0.6$. A small gas fraction and a high SFR in spiral galaxies suggest that the gravitational instabilities efficiently convert the gas into stars depleting the gas reservoir. On the other hand, higher gas reserves in irregular galaxies suggest that the gravitational instabilities are inefficient at converting the gas into stars. In order to get a better picture of how $Q^{Min}_{RW}$ is connected with $f^{Gas}$ and the SFR, we compute the time scale for the growth of gravitational instabilities, which measures how quickly the gas is converted into stars. 
The time scale for the growth of gravitational instabilities is given by \citep{talbor1975evolution, leroy2008star, wong2009timescale}
\begin{equation}
    \tau = \frac{2 \pi}{\omega_{J}}.
\end{equation}
In the above equation, $\omega_{J}$ is defined as the growth rate of gravitational instabilities  and is given as 
\begin{equation}
\omega_{J}=\frac{\pi G \Sigma_{Gas}}{\sigma_{Gas}} ( 1 + \frac{\Sigma_{Stars} \sigma_{Gas} }{\Sigma_{Gas} \sigma_{Stars}}).
\end{equation}

In Figure 8, we show the time scale for the growth of gravitational instabilities as a function of gas fraction [panel - 1], star formation rate [panel -2 ], and the two-component stability parameter [panel - 3]. From the first panel, we see that the galaxies in which gravitational instabilities persist for a more extended period have a higher gas fraction. On the other hand, the galaxies that sustain instabilities for a short time have a lower gas fraction. Further, from the second and the third panel, we see that galaxies in which gravitational instabilities persist for an extended time have lower star formation rates and a higher value of $Q^{Min}_{RW}$. 
Thus, instabilities characterized by $Q_{RW}$ close to marginal stability acting for short time scales are responsible for depleting the gas reservoirs. On the other hand, instabilities persisting for a longer period in galaxies characterized by a higher $Q_{RW}$ have a larger gas fraction and a lower star formation rate. The median value of $\tau$  for the irregular galaxies is equal to $214\, Myr$, and $\tau$ for the spiral galaxies is equal to $104\, Myr$. The values of $\tau$ obtained for the nearby spiral galaxies in this study are consistent with the values of $\tau$ obtained by \cite{wong2009timescale} for 16 galaxies taken from THINGS sample $(\tau = 100 \,Myr)$. The gravitational instabilities act over short times scales in spiral galaxies $(\tau=104\, Myr)$  and efficiently convert the available gas into stars, which explains why the spiral galaxies have a small gas fraction $(f^{Gas}=0.1)$ and a high star formation rate $(SFR =0.2M_{\odot}yr^{-1})$.
Whereas on the other hand, in irregular galaxies, the gravitational instabilities persist for a longer time $(214\, Myr)$ and convert the gas into stars more gradually, which explains why irregular galaxies have a higher gas fraction $(f^{Gas}=0.6)$ and a lower star formation rate compared to the spiral galaxies. Finally, we note that the gravitational instabilities persist in star-forming galaxies at $z = 4.5$ for only $6 Myr$, and the median gas fraction is equal to $0.5$. A small value of $Q_{RW}$, a high star formation rate, and a large gas fraction
indicate that the galaxies at $z=4.5$ are in an active star-forming stage, unlike nearby galaxies, which have possibly reached their threshold gas fraction and star formation rate. It suggests that spiral galaxies possibly 
undergo intense star formation in multiple short bursts before exhausting their gas reserves and reach threshold star formation rate and stability level. Also, unlike spiral galaxies, irregular galaxies convert gas into stars more gradually over a much longer time scale compared to spiral galaxies. The mechanism uniting the instability levels and the star formation rates with the gas fraction are consistent with the recent results from the FIRE -2 simulations \citep{hopkins2018fire}. Using the FIRE - 2 simulations \cite{parul2023imprint, flores2021time} show that star formation in Milky Way-like galaxies occurs in short bursts at high redshifts and proceeds more 
steadily at low redshifts. The results obtained in this work suggest a simple mechanism in which galaxies characterized by $Q_{RW}$ close to marginal stability levels undergo intense star formation activity for a short time scale, depleting the gas reserves. Whereas in the second scenario, the star formation proceeds more slowly over longer time scales in galaxies with a relatively higher value of $Q_{RW}$, gradually converting the available gas into stars. 

\section{Summary}
In this work, we have studied the stability of the nearby galaxies, which are a part of the SPARC catalog, using the two-component stability criterion proposed by \cite{romeo2011effective}. We find: 

\begin{enumerate}

\item The net stability in the galaxies are primarily driven by the stellar disc.

\item Despite their diverse morphological properties, $91\%$ galaxies in the SPARC galaxy catalog have a $Q^{Min}_{RW}>1$, indicating stability against axisymmetric instabilities. Further, atleast $50\%$ of the galaxies have $Q^{Min}_{RW}$ between $2-3$, indicating critical stability against non-axisymmetric instabilities and gas dissipation.

\item The galaxies with $Q^{Min}_{RW}<1$ have a median star formation rate equal to $0.4$ $M_{\odot}yr^{-1}$ which is higher than  galaxies with $Q^{Min}_{RW}>1$ ($SFR = 0.07 M_{\odot}yr^{-1}$). Further, we find that the median star formation rates in the spiral galaxies $(SFR= 0.2M_{\odot}yr^{-1})$ is higher than the irregular galaxies $(SFR= 0.02M_{\odot}yr^{-1})$. We note that galaxies with higher star formation rates and lower stability parameters have a lower gas fraction.

\item The stellar disc attains minimum stability close to the center, at $1.4R_{D}$, whereas the gas disc attains minimum values at $5.1R_{D}$. However, the net stability of the galaxies determined by the two-component stability parameter $Q_{RW}$ attains a minimum value at an intermediate value equal to $2.8R_{D}$. 
 
\item Finally, we compare the stability levels of the nearby galaxies in SPARC catalog with dynamically cold disc galaxies observed at $z \,=\, 4.5$. We find that the galaxies at $z\,=\,4.5$ are characterized by low stability and a high star formation rate compared with the nearby galaxies. 

\item We find that galaxies with $Q^{Min}_{RW}<1$ in the SPARC catalogue have a median $SFR \,=\, 0.4 M_{\odot} yr^{-1}$, whereas galaxies with $Q^{Min}_{RW}<1$ at $z\,=\,4.5$ have a median SFR equal to $\rm 7\times10^{2} M_{\odot} yr^{-1}$.  This indicates that the gravitational instabilities in the galaxies can self-regulate the stability levels. Also, the galaxies at $z=4.5$ have $f^{Gas} = 0.5$ and nearby galaxies have a median $f^{Gas} = 0.1$. This suggests that the galaxies at $z=4.5$ are in an active star-forming stage, whereas the nearby galaxies have reached a threshold star formation level and gas fraction.

\item In order to better understand how the galaxies can self-regulate the stability levels, we derive the stability of the galaxies at $z\,=\,4.5$ and those in the SPARC catalog by neglecting the contribution of the dark matter to the total potential. The stability levels of galaxies at $z\,=\,4.5$ are unchanged upon eliminating the dark matter's contribution from the total potential. On the other hand, the global median of $Q_{RW}$ changes from $3.0$ to $2.3$ upon eliminating dark matter's contribution from total potential in the case of nearby galaxies. The galaxies in the SPARC catalog remain stable against axisymmetric instabilities upon removing the dark matter's contribution from the total potential suggesting  that the baryons can regulate the stability levels, atleast statistically.

\item In this study, we find that the galaxies which have a higher value of $Q_{RW}$ typically also have a higher gas fraction and a lower star formation rate. So, to understand how the star formation rate and the two-component parameter are related to the gas fraction, we measure the time scale for the growth of gravitational instabilities, which measures how quickly gas is converted into stars. We find that gravitational instabilities act for a short period in spiral galaxies, efficiently converting the gas into stars, which explains why spirals have relatively smaller gas fraction than irregular galaxies. Whereas gravitational instabilities persist over a more extended period and gradually convert the available gas into stars in the irregular galaxies.

\end{enumerate}

\section{Acknowledgement}
Aditya would like to thank the referee for their insightful comments that improved the quality of this manuscript.
The work uses data from Spitzer Photometry and \& Accurate Rotation Curves (SPARC) database \citep{lelli2016sparc}. 
Aditya is supported by a DST-SERB grant [CRG/2021/005174].

\section{Data Availability}
The data used in this study is publicly available at \textcolor{blue}{http://astroweb.cwru.edu/SPARC/}.

\newpage
\small{\bibliographystyle{mnras}}
\bibliography{ms} 

\begin{thebibliography}{}
\makeatletter
\relax
\def\mn@urlcharsother{\let\do\@makeother \do\$\do\&\do\#\do\^\do\_\do\%\do\~}
\def\mn@doi{\begingroup\mn@urlcharsother \@ifnextchar [ {\mn@doi@}
  {\mn@doi@[]}}
\def\mn@doi@[#1]#2{\def\@tempa{#1}\ifx\@tempa\@empty \href
  {http://dx.doi.org/#2} {doi:#2}\else \href {http://dx.doi.org/#2} {#1}\fi
  \endgroup}
\def\mn@eprint#1#2{\mn@eprint@#1:#2::\@nil}
\def\mn@eprint@arXiv#1{\href {http://arxiv.org/abs/#1} {{\tt arXiv:#1}}}
\def\mn@eprint@dblp#1{\href {http://dblp.uni-trier.de/rec/bibtex/#1.xml}
  {dblp:#1}}
\def\mn@eprint@#1:#2:#3:#4\@nil{\def\@tempa {#1}\def\@tempb {#2}\def\@tempc
  {#3}\ifx \@tempc \@empty \let \@tempc \@tempb \let \@tempb \@tempa \fi \ifx
  \@tempb \@empty \def\@tempb {arXiv}\fi \@ifundefined
  {mn@eprint@\@tempb}{\@tempb:\@tempc}{\expandafter \expandafter \csname
  mn@eprint@\@tempb\endcsname \expandafter{\@tempc}}}

\bibitem[\protect\citeauthoryear{Agertz, Romeo  \& Grisdale}{Agertz
  et~al.}{2015}]{agertz2015characterizing}
Agertz O.,  Romeo A.~B.,   Grisdale K.,  2015, Monthly Notices of the Royal
  Astronomical Society, 449, 2156

\bibitem[\protect\citeauthoryear{Bacchini, Fraternali, Iorio, Pezzulli, Marasco
   \& Nipoti}{Bacchini et~al.}{2020}]{bacchini2020evidence}
Bacchini C.,  Fraternali F.,  Iorio G.,  Pezzulli G.,  Marasco A.,   Nipoti C.,
   2020, Astronomy \& Astrophysics, 641, A70

\bibitem[\protect\citeauthoryear{Barnes \& Hernquist}{Barnes \&
  Hernquist}{1992}]{barnes1992formation}
Barnes J.~E.,  Hernquist L.,  1992, Nature, 360, 715

\bibitem[\protect\citeauthoryear{Binney \& Tremaine}{Binney \&
  Tremaine}{2011}]{binney2011galactic}
Binney J.,  Tremaine S.,  2011, Galactic dynamics.
 Vol. 13, Princeton university press

\bibitem[\protect\citeauthoryear{Boissier, Prantzos, Boselli  \&
  Gavazzi}{Boissier et~al.}{2003}]{boissier2003star}
Boissier S.,  Prantzos N.,  Boselli A.,   Gavazzi G.,  2003, Monthly Notices of
  the Royal Astronomical Society, 346, 1215

\bibitem[\protect\citeauthoryear{Bournaud, Jog  \& Combes}{Bournaud
  et~al.}{2007}]{bournaud2007multiple}
Bournaud F.,  Jog C.~J.,   Combes F.,  2007, Astronomy \& Astrophysics, 476,
  1179

\bibitem[\protect\citeauthoryear{Brunetti, Wilson, Sliwa, Schinnerer, Aalto  \&
  Peck}{Brunetti et~al.}{2021}]{brunetti2021highly}
Brunetti N.,  Wilson C.~D.,  Sliwa K.,  Schinnerer E.,  Aalto S.,   Peck A.~B.,
   2021, Monthly Notices of the Royal Astronomical Society, 500, 4730

\bibitem[\protect\citeauthoryear{Burkert \& Hartmann}{Burkert \&
  Hartmann}{2013}]{burkert2013dependence}
Burkert A.,  Hartmann L.,  2013, The Astrophysical Journal, 773, 48

\bibitem[\protect\citeauthoryear{Dawson, McClure-Griffiths, Wong, Dickey,
  Hughes, Fukui  \& Kawamura}{Dawson et~al.}{2013}]{dawson2013supergiant}
Dawson J.,  McClure-Griffiths N.,  Wong T.,  Dickey J.~M.,  Hughes A.,  Fukui
  Y.,   Kawamura A.,  2013, The Astrophysical Journal, 763, 56

\bibitem[\protect\citeauthoryear{Elmegreen}{Elmegreen}{2011}]{elmegreen2011gravitational}
Elmegreen B.~G.,  2011, The Astrophysical Journal, 737, 10

\bibitem[\protect\citeauthoryear{Flores~Vel{\'a}zquez
  et~al.,}{Flores~Vel{\'a}zquez et~al.}{2021}]{flores2021time}
Flores~Vel{\'a}zquez J.~A.,  et~al., 2021, Monthly Notices of the Royal
  Astronomical Society, 501, 4812

\bibitem[\protect\citeauthoryear{Garg \& Banerjee}{Garg \&
  Banerjee}{2017}]{garg2017origin}
Garg P.,  Banerjee A.,  2017, Monthly Notices of the Royal Astronomical
  Society, 472, 166

\bibitem[\protect\citeauthoryear{Genzel et~al.,}{Genzel
  et~al.}{2017}]{genzel2017strongly}
Genzel R.,  et~al., 2017, Nature, 543, 397

\bibitem[\protect\citeauthoryear{Goldreich \& Lynden-Bell}{Goldreich \&
  Lynden-Bell}{1965}]{goldreich1965gravitational}
Goldreich P.,  Lynden-Bell D.,  1965, Monthly Notices of the Royal Astronomical
  Society, 130, 97

\bibitem[\protect\citeauthoryear{Griv \& Gedalin}{Griv \&
  Gedalin}{2012}]{griv2012stability}
Griv E.,  Gedalin M.,  2012, Monthly Notices of the Royal Astronomical Society,
  422, 600

\bibitem[\protect\citeauthoryear{Gruendl \& Chu}{Gruendl \&
  Chu}{2009}]{gruendl2009high}
Gruendl R.~A.,  Chu Y.-H.,  2009, The Astrophysical Journal Supplement Series,
  184, 172

\bibitem[\protect\citeauthoryear{Hoffmann \& Romeo}{Hoffmann \&
  Romeo}{2012}]{hoffmann2012effect}
Hoffmann V.,  Romeo A.~B.,  2012, Monthly Notices of the Royal Astronomical
  Society, 425, 1511

\bibitem[\protect\citeauthoryear{Hopkins et~al.,}{Hopkins
  et~al.}{2018}]{hopkins2018fire}
Hopkins P.~F.,  et~al., 2018, Monthly Notices of the Royal Astronomical
  Society, 480, 800

\bibitem[\protect\citeauthoryear{Jog}{Jog}{1996}]{jog1996local}
Jog C.~J.,  1996, Monthly Notices of the Royal Astronomical Society, 278, 209

\bibitem[\protect\citeauthoryear{Kennicutt~Jr}{Kennicutt~Jr}{1998}]{kennicutt1998star}
Kennicutt~Jr R.~C.,  1998, Annual Review of Astronomy and Astrophysics, 36, 189

\bibitem[\protect\citeauthoryear{Lelli, McGaugh  \& Schombert}{Lelli
  et~al.}{2016}]{lelli2016sparc}
Lelli F.,  McGaugh S.~S.,   Schombert J.~M.,  2016, The Astronomical Journal,
  152, 157

\bibitem[\protect\citeauthoryear{Leroy, Walter, Brinks, Bigiel, De~Blok, Madore
   \& Thornley}{Leroy et~al.}{2008}]{leroy2008star}
Leroy A.~K.,  Walter F.,  Brinks E.,  Bigiel F.,  De~Blok W.,  Madore B.,
  Thornley M.,  2008, The astronomical journal, 136, 2782

\bibitem[\protect\citeauthoryear{Leroy et~al.,}{Leroy
  et~al.}{2016}]{leroy2016portrait}
Leroy A.~K.,  et~al., 2016, The Astrophysical Journal, 831, 16

\bibitem[\protect\citeauthoryear{Leroy et~al.,}{Leroy
  et~al.}{2021}]{leroy2021phangs}
Leroy A.~K.,  et~al., 2021, The Astrophysical Journal Supplement Series, 255,
  19

\bibitem[\protect\citeauthoryear{Li, Mac~Low  \& Klessen}{Li
  et~al.}{2005}]{li2005control}
Li Y.,  Mac~Low M.-M.,   Klessen R.~S.,  2005, The Astrophysical Journal, 620,
  L19

\bibitem[\protect\citeauthoryear{Lin, Papaloizou, Levy  \& Lunine}{Lin
  et~al.}{1993}]{lin1993protostars}
Lin D.,  Papaloizou J.,  Levy E.,   Lunine J.,  1993, University of Arizona,
  Tucson, p.~749

\bibitem[\protect\citeauthoryear{Marasco, Fraternali, Van~der Hulst  \&
  Oosterloo}{Marasco et~al.}{2017}]{marasco2017distribution}
Marasco A.,  Fraternali F.,  Van~der Hulst J.,   Oosterloo T.,  2017, Astronomy
  \& Astrophysics, 607, A106

\bibitem[\protect\citeauthoryear{Martig, Bournaud, Teyssier  \& Dekel}{Martig
  et~al.}{2009}]{martig2009morphological}
Martig M.,  Bournaud F.,  Teyssier R.,   Dekel A.,  2009, The Astrophysical
  Journal, 707, 250

\bibitem[\protect\citeauthoryear{McGaugh \& Schombert}{McGaugh \&
  Schombert}{2014}]{mcgaugh2014color}
McGaugh S.~S.,  Schombert J.~M.,  2014, The Astronomical Journal, 148, 77

\bibitem[\protect\citeauthoryear{Meidt}{Meidt}{2022}]{meidt2022molecular}
Meidt S.~E.,  2022, The Astrophysical Journal, 937, 88

\bibitem[\protect\citeauthoryear{Meidt et~al.,}{Meidt
  et~al.}{2023}]{meidt2023phangs}
Meidt S.~E.,  et~al., 2023, The Astrophysical Journal Letters, 944, L18

\bibitem[\protect\citeauthoryear{M{\'e}ra, Mizony  \& Baillon}{M{\'e}ra
  et~al.}{1996}]{mera1996disk}
M{\'e}ra D.,  Mizony M.,   Baillon J.,  1996, Preprint (submitted to Astron.
  Astrophys., and Mon. Not. R. Astron. Soc.), 97

\bibitem[\protect\citeauthoryear{Mestel}{Mestel}{1963}]{mestel1963galactic}
Mestel L.,  1963, Monthly Notices of the Royal Astronomical Society, 126, 553

\bibitem[\protect\citeauthoryear{Mihos \& Hernquist}{Mihos \&
  Hernquist}{1994}]{mihos1994triggering}
Mihos J.~C.,  Hernquist L.,  1994, The Astrophysical Journal, 425, L13

\bibitem[\protect\citeauthoryear{Mihos, Walker, Hernquist, de Oliveira  \&
  Bolte}{Mihos et~al.}{1995}]{mihos1995merger}
Mihos J.~C.,  Walker I.~R.,  Hernquist L.,  de Oliveira C.~M.,   Bolte M.,
  1995, The Astrophysical Journal, 447, L87

\bibitem[\protect\citeauthoryear{Mogotsi, de Blok, Cald{\'u}-Primo, Walter,
  Ianjamasimanana  \& Leroy}{Mogotsi et~al.}{2016}]{mogotsi2016hi}
Mogotsi K.,  de Blok W.,  Cald{\'u}-Primo A.,  Walter F.,  Ianjamasimanana R.,
   Leroy A.,  2016, The Astronomical Journal, 151, 15

\bibitem[\protect\citeauthoryear{Narayanan \& Banerjee}{Narayanan \&
  Banerjee}{2022}]{narayanan2022superthin}
Narayanan G.,  Banerjee A.,  2022, Monthly Notices of the Royal Astronomical
  Society, 514, 5126

\bibitem[\protect\citeauthoryear{Pandey \& Van De~Bruck}{Pandey \& Van
  De~Bruck}{1999}]{pandey1999gravitational}
Pandey U.,  Van De~Bruck C.,  1999, Monthly Notices of the Royal Astronomical
  Society, 306, 181

\bibitem[\protect\citeauthoryear{Parul, Bailin, Wetzel, Gurvich,
  Faucher-Gigu{\`e}re, Hafen, Stern  \& Snaith}{Parul
  et~al.}{2023}]{parul2023imprint}
Parul H.,  Bailin J.,  Wetzel A.,  Gurvich A.~B.,  Faucher-Gigu{\`e}re C.-A.,
  Hafen Z.,  Stern J.,   Snaith O.,  2023, arXiv preprint arXiv:2301.07692

\bibitem[\protect\citeauthoryear{Pedrosa, Tissera  \& De~Rossi}{Pedrosa
  et~al.}{2014}]{pedrosa2014morphology}
Pedrosa S.~E.,  Tissera P.~B.,   De~Rossi M.~E.,  2014, Astronomy \&
  Astrophysics, 567, A47

\bibitem[\protect\citeauthoryear{Rafikov}{Rafikov}{2001}]{rafikov2001local}
Rafikov R.~R.,  2001, Monthly Notices of the Royal Astronomical Society, 323,
  445

\bibitem[\protect\citeauthoryear{Renaud et~al.,}{Renaud
  et~al.}{2018}]{renaud2018morphology}
Renaud F.,  et~al., 2018, Monthly Notices of the Royal Astronomical Society,
  473, 585

\bibitem[\protect\citeauthoryear{Rizzo, Vegetti, Powell, Fraternali, McKean,
  Stacey  \& White}{Rizzo et~al.}{2020}]{rizzo2020dynamically}
Rizzo F.,  Vegetti S.,  Powell D.,  Fraternali F.,  McKean J.,  Stacey H.,
  White S.,  2020, Nature, 584, 201

\bibitem[\protect\citeauthoryear{Rizzo, Vegetti, Fraternali, Stacey  \&
  Powell}{Rizzo et~al.}{2021}]{rizzo2021dynamical}
Rizzo F.,  Vegetti S.,  Fraternali F.,  Stacey H.~R.,   Powell D.,  2021,
  Monthly Notices of the Royal Astronomical Society, 507, 3952

\bibitem[\protect\citeauthoryear{Romeo \& Falstad}{Romeo \&
  Falstad}{2013}]{romeo2013simple}
Romeo A.~B.,  Falstad N.,  2013, Monthly Notices of the Royal Astronomical
  Society, 433, 1389

\bibitem[\protect\citeauthoryear{Romeo \& Mogotsi}{Romeo \&
  Mogotsi}{2017}]{romeo2017drives}
Romeo A.~B.,  Mogotsi K.~M.,  2017, Monthly Notices of the Royal Astronomical
  Society, 469, 286

\bibitem[\protect\citeauthoryear{Romeo \& Mogotsi}{Romeo \&
  Mogotsi}{2018}]{romeo2018angular}
Romeo A.~B.,  Mogotsi K.~M.,  2018, Monthly Notices of the Royal Astronomical
  Society: Letters, 480, L23

\bibitem[\protect\citeauthoryear{Romeo \& Wiegert}{Romeo \&
  Wiegert}{2011}]{romeo2011effective}
Romeo A.~B.,  Wiegert J.,  2011, Monthly Notices of the Royal Astronomical
  Society, 416, 1191

\bibitem[\protect\citeauthoryear{Romeo, Agertz  \& Renaud}{Romeo
  et~al.}{2020}]{romeo2020lenticulars}
Romeo A.~B.,  Agertz O.,   Renaud F.,  2020, Monthly Notices of the Royal
  Astronomical Society, 499, 5656

\bibitem[\protect\citeauthoryear{Scannapieco, White, Springel  \&
  Tissera}{Scannapieco et~al.}{2009}]{scannapieco2009formation}
Scannapieco C.,  White S.~D.,  Springel V.,   Tissera P.~B.,  2009, Monthly
  Notices of the Royal Astronomical Society, 396, 696

\bibitem[\protect\citeauthoryear{Talbor~Jr \& Arnett}{Talbor~Jr \&
  Arnett}{1975}]{talbor1975evolution}
Talbor~Jr R.,  Arnett W.,  1975, Astrophysical Journal, 197, 551

\bibitem[\protect\citeauthoryear{Tamburro, Rix, Leroy, Mac~Low, Walter,
  Kennicutt, Brinks  \& De~Blok}{Tamburro et~al.}{2009}]{tamburro2009driving}
Tamburro D.,  Rix H.-W.,  Leroy A.,  Mac~Low M.-M.,  Walter F.,  Kennicutt R.,
  Brinks E.,   De~Blok W.,  2009, The Astronomical Journal, 137, 4424

\bibitem[\protect\citeauthoryear{Toomre}{Toomre}{1964}]{toomre1964gravitational}
Toomre A.,  1964, The Astrophysical Journal, 139, 1217

\bibitem[\protect\citeauthoryear{Toomre \& Toomre}{Toomre \&
  Toomre}{1972}]{toomre1972galactic}
Toomre A.,  Toomre J.,  1972, The Astrophysical Journal, 178, 623

\bibitem[\protect\citeauthoryear{Villanueva et~al.,}{Villanueva
  et~al.}{2021}]{villanueva2021edge}
Villanueva V.,  et~al., 2021, The Astrophysical Journal, 923, 60

\bibitem[\protect\citeauthoryear{Walter, Brinks, De~Blok, Bigiel, Kennicutt,
  Thornley  \& Leroy}{Walter et~al.}{2008}]{walter2008things}
Walter F.,  Brinks E.,  De~Blok W.,  Bigiel F.,  Kennicutt R.~C.,  Thornley
  M.~D.,   Leroy A.,  2008, The Astronomical Journal, 136, 2563

\bibitem[\protect\citeauthoryear{Wang \& Silk}{Wang \&
  Silk}{1994}]{wang1994gravitational}
Wang B.,  Silk J.,  1994, The Astrophysical Journal, 427, 759

\bibitem[\protect\citeauthoryear{Westfall, Andersen, Bershady, Martinsson,
  Swaters  \& Verheijen}{Westfall et~al.}{2014}]{westfall2014diskmass}
Westfall K.~B.,  Andersen D.~R.,  Bershady M.~A.,  Martinsson T.~P.,  Swaters
  R.~A.,   Verheijen M.~A.,  2014, The Astrophysical Journal, 785, 43

\bibitem[\protect\citeauthoryear{Wong}{Wong}{2009}]{wong2009timescale}
Wong T.,  2009, The Astrophysical Journal, 705, 650

\makeatother
\end{thebibliography}
\end{document}